\documentclass[%
preprint,
 amsmath,amssymb,
 aps,
]{revtex4-1}

\usepackage{color}
\usepackage{graphicx}% Include figure files
\usepackage{dcolumn}% Align table columns on decimal point
\usepackage{bm}% bold math
\def\sub#1{_{\mathrm{#1}}}
\def\up#1{^{\mathrm{#1}}}
\def\Vec#1{\boldsymbol #1}

\newcommand {\beq}{\begin{eqnarray}}
\newcommand {\eeq}{\end{eqnarray}}

\begin{document}

%\preprint{APS/123-QED}

\title{
Non-relativistic Nambu-Goldstone modes associated with  spontaneously broken space-time and internal symmetries  
}% Force line breaks with \\
%\thanks{Geometrical determination of internal degrees of freedom localized in vortex cores}%

\author{Michikazu Kobayashi$^1$, Muneto Nitta$^2$}
\affiliation{%
$^1$Department of Physics, Kyoto University, Oiwake-cho, Kitashirakawa, Sakyo-ku, Kyoto 606-8502, Japan, \\
$^2$Department of Physics, and Research and Education Center for Natural
Sciences, Keio University, Hiyoshi 4-1-1, Yokohama, Kanagawa 223-8521, Japan
}%

\date{\today}% It is always \today, today,

\begin{abstract}
We show that 
a momentum operator of a translational symmetry 
may not commute with 
an internal symmetry operator
in the presence of a topological soliton 
in non-relativistic
theories. 
As a striking consequence, there appears 
a coupled Nambu-Goldstone mode 
with a quadratic dispersion 
consisting of translational and internal zero modes 
in the vicinity of a domain wall 
in an $O(3)$ sigma model, 
a magnetic domain wall in ferromagnets with an easy axis.
  
\end{abstract}


\maketitle

\section{Introduction}
Symmetry is one of the most important guiding principles in describing nature in quantum physics. 
In particular, relativistic quantum field theories rely on 
symmetry principle and have been quite successful in
unifying fundamental forces.
Gauge symmetries of 
electromagnetic, weak and strong 
interactions can be unified to 
one group in Grand unified theories.
However,  
space-time symmetry related to gravity 
could not be, 
because of the Coleman-Mandula theorem \cite{Coleman:1967ad} 
showing that internal symmetry and space-time symmetry 
must be a direct product \cite{footnote1}.
These are all based on relativistic field theories. 

In this Letter, we find a symmetry algebra including both the space-time symmetry and an internal symmetry 
in
{\it non-relativistic}
field theory:
\begin{align}
 \left[ P, \Theta \right] = W \neq 0  \label{eq:alg}
\end{align}
where $P$ is a translational operator, 
$\Theta$ is an internal symmetry operator and
$W$ is a ``central''  extension. 
We show that the commutation relation, Eq.~(\ref{eq:alg}), is possible 
in the presence of a topological soliton 
in non-relativistic
theories.  
In a practical model, the central charge $W$ 
is a topological charge of a domain wall \cite{footnote2}. 
The central charge $W$ vanishes in the corresponding relativistic model. 
As a consequence of our novel algebra, 
a Nambu-Goldstone (NG) mode 
for the translational symmetry 
(a ripple mode or ripplon if quantized) 
is coupled to that for the internal $U(1)$ symmetry 
(a magnon)
when the presence of 
a domain wall breaks both the symmetries;
They give rise to one NG mode with a quadratic dispersion relation,
although two symmetry generators are spontaneously broken.
This is in contrast to the corresponding relativistic  model, 
in which these two modes, ripplon and magnon, appear independently 
with linear dispersion relations, 
which corresponds to the fact that $W$ vanishes. 
This phenomenon itself is already known as type-II 
or type-B
NG modes in non-relativistic theories \cite{Nielsen:1975hm,Nambu:2004yia,
Watanabe:2011ec,Watanabe:2012hr,Hidaka:2012ym}. 
However, previously known examples are either NG modes of 
both internal symmetries such as ferromagnets 
or those of both space-time 
symmetry such as vortices and lumps (skyrmions) \cite{Kobayashi:2013gba,Watanabe:2014pea},
while our case mixes them together because of 
Eq.~(\ref{eq:alg}).
We derive the dispersion relation in two approaches; 
the effective field theory for topological solitons 
and Bogoliubov theory. 
We also study a model interpolating between 
relativistic and non-relativistic theories and 
find unexpectedly a coupled NG mode in an interpolating region even though it has the Lorentz invariance.

%\newpage

\section{Models and a domain wall}\label{sec:model}

We start from the following relativistic, and non-relativistic $\mathbb{C}P^1$ Lagrangian densities $\mathcal{L}\sub{rel}$ and $\mathcal{L}\sub{nrel}$ with an Ising-type potential:
\begin{align}
\mathcal{L}\sub{rel} = \frac{|\dot{u}|^2 - |\nabla u|^2 - m^2 |u|^2}{(1 + |u|^2)^2}, \quad
\mathcal{L}\sub{nrel} = \frac{i ( u^\ast \dot{u} - \dot{u}^\ast u)}{2 (1 + |u|^2)} - \frac{|\nabla u|^2 + m^2 |u|^2}{(1 + |u|^2)^2}, \label{eq-Lagrangian}
\end{align}
where $u \in \mathbb{C}$ is the complex projective coordinate defined as $\phi^T = (1, u)^T / \sqrt{1 + |u|^2}$ with 
normalized two scalar fields $\phi = (\phi_1, \phi_2)^T$.
$\mathcal{L}\sub{rel}$ and $\mathcal{L}\sub{nrel}$ are equivalent to $O(3)$ nonlinear sigma models:
\begin{align}
\begin{split}
& \mathcal{L}\sub{rel} = \frac{1}{4} \{ |\dot{\Vec{n}}|^2 - |\nabla \Vec{n}|^2 - m^2(1 - n_3^2) \}, \\
& \mathcal{L}\sub{nrel} = \frac{\dot{n}_1 n_2 - n_1 \dot{n}_2}{2 (1 + n_3)} - \frac{1}{4} \{ |\nabla \Vec{n}|^2 + m^2(1 - n_3^2) \}, \label{eq-sigma}
\end{split}
\end{align}
under the Hopf map for a three-vector of scalar fields $\Vec{n} \equiv \phi^\dagger  \Vec{\sigma}  \phi $ with the Pauli matrices $\Vec{\sigma}$.
These models describe ferromagnets with one easy axis.

A Lagrangian density interpolating between $\mathcal{L}\sub{rel}$ and $\mathcal{L}\sub{nrel}$ is given as the following form:
\begin{align}
\begin{split}
\mathcal{L}\sub{G} &= \phi_0^2 \bigg\{ \frac{|\dot{u}|^2}{c^2 (1 + |u|^2)^2} - \frac{|\nabla u|^2}{(1 + |u|^2)^2} + \frac{i M ( u^\ast \dot{u} - \dot{u}^\ast u )}{\hbar(1 + |u|^2)} - \frac{m^2 |u|^2}{(1 + |u|^2)^2} \bigg\}, \label{eq-gauged-Lagrangian}
\end{split}
\end{align}
where $\phi_0^2 (> 0)$ is a real positive (decay) constant having the dimension of $[\mathrm{energy}] / [\mathrm{length}]$.
In the following, we omit $\phi_0^2$ by measuring $\mathcal{L}\sub{G}$ in the unit of $\phi_0^2$: $\mathcal{L}\sub{G} \to \phi_0^2 \mathcal{L}\sub{G}$.
A detailed derivation of the Lagrangian density $\mathcal{L}\sub{G}$ is discussed in 
Appendix \ref{sec:rel-nrel}. 
$\mathcal{L}\sub{rel}$ is obtained as the massless limit $M \to 0$ of $\mathcal{L}\sub{G}$, while $\mathcal{L}\sub{nrel}$ is the nonrelativistic limit $c \to \infty$ of $\mathcal{L}\sub{G}$ \cite{ultrarelativistic}.

The actions $S = \int d^4x\: \mathcal{L}\sub{G}$ is invariant under a global discrete $\mathbb{Z}_2$ transformation: $u \leftrightarrow 1 / u^\ast$, a global $U(1)$ phase rotation: $u \to u e^{i \alpha}$.
$S$ is also invariant under the Poincar\'e transformation as long as $c$ is positive finite, or the Galilean transformation in the non-relativistic limit $c \to \infty$, 
as shown in Appendix~\ref{sec:rel-nrel}.
There are two discrete vacua $|u| = 0$ or $|u| \to \infty$, and $m > 0$ is defined as the energy gap between them.
For these vacua, the $\mathbb{Z}_2$ symmetry for the global discrete transformation is spontaneously broken.
In the framework of the nonlinear sigma models \eqref{eq-sigma}, the vacua are expressed as $n_1 = n_2 = 0$, $n_3 = \pm 1$, and the last $m^2 (1 - n_3^2)$ terms are regarded as the Ising potential.

Dynamics of $u$ can be obtained by the Euler-Lagrange equation for $\mathcal{L}\sub{G}$:
\begin{align}
\begin{split}
& \frac{(1 + |u|^2) \ddot{u} - 2 u^\ast \dot{u}^2}{c^2} - \frac{2 i M (1 + |u|^2) \dot{u}}{\hbar} = (1 + |u|^2) \nabla^2 u - 2 u^\ast (\nabla u)^2 - m^2 ( 1 - |u|^2) u. \label{eq-dynamics}
\end{split}
\end{align}

We next consider a static domain or anti-domain wall solution interpolating the two vacua.
The flat and static domain wall solution perpendicular to the $z$-axis is \cite{Abraham:1992vb} 
(see Appendix \ref{sec:topological})
\begin{align}
u_0 = \exp\{ m (z - Z) + i \alpha\}, \label{eq-domain-wall}
\end{align}
where $\alpha$ ($0 \leq \alpha < 2 \pi$) and $Z \in \mathbb{R}$ are phase and translational moduli of the domain wall. 
This is known as a magnetic domain wall in ferromagnets 
with an easy axis. 

In the presence of the domain wall, the $H_1 \simeq U(1) \times \mathbb{R}^3$ symmetry is further spontaneously broken, where $U(1)$ is the global symmetry for the internal phase rotation and $\mathbb{R}^3$ is the three-dimensional translational symmetry in a space.
The remaining symmetry is $H_2 \simeq \mathbb{R}^2_{xy}$, where $\mathbb{R}^2_{xy}$ indicates the translation along the $xy$--plane \cite{rotational-symmetry}.
Breaking symmetries $H_1 / H_2 \simeq U(1) \times \mathbb{R}_z$ due to the domain wall are the internal $U(1)$ phase rotation and translation along the $z$ direction, and two moduli $\alpha$ and $Z$ in Eq. \eqref{eq-domain-wall} are regarded as corresponding NG modes in the vicinity of the domain wall.
The NG mode $\alpha$ is the phase mode known as a magnon localized in the domain wall.
The other NG mode for $Z$ is the translational surface mode of the domain wall, known as a ripple mode, or ripplon if quantized, 
in condensed matter physics.
In the following, we show that the localized magnon and ripplon are coupled to each other to become one ``coupled ripplon" mode with fixed dispersion relation and amplitude.

%%%%%%%%%%%%%%%%
\section{Low-energy effective theory of a domain wall}
We next consider the NG modes excited along the domain wall 
by constructing the effective theory on a domain wall 
by the moduli approximation \cite{Manton:1981mp}. 
Introducing $\Vec{r} = (x, y)$, and $t$ dependences of two moduli $\alpha$ and $Z$ as $\alpha(\Vec{r},t)$ and $Z(\Vec{r},t)$, we consider the ansatz $u$ as
\begin{align}
u = \exp[ m \{z - Z(\Vec{r},t)\} + i \alpha(\Vec{r},t)]. \label{eq-ripplon-ansatz}
\end{align}
Inserting Eq. \eqref{eq-ripplon-ansatz} to Eq. \eqref{eq-gauged-Lagrangian}, the effective Lagrangian $L\up{eff}\sub{G}$ defined as $L\up{eff}\sub{G} = \lim_{L \to \infty} \int_{-L}^L dz\: \mathcal{L}\sub{G}$ becomes
\begin{align}
\begin{split}
L\up{eff}\sub{G} = \frac{m^2 (\dot{Z}^2 / c^2 - |\nabla_{\Vec{r}} Z|^2) + \dot{\alpha}^2 / c^2 - |\nabla_{\Vec{r}} \alpha|^2}{2 m} + \frac{2 M (Z - L) \dot{\alpha}}{\hbar} - m + O(\nabla^3),
\end{split}
\end{align}
up to the quadratic order in derivatives. 
Here, $\nabla_{\Vec{r}} = (\partial_x ,\partial_y)$ 
is the derivative in the $xy$--plane.
The constant term $m$ is the tension (the energy per unit area) of the static flat domain wall.
The case in the massless limit $M \to 0$ was already obtained before \cite{Arai:2002xa}.

The low-energy dynamics of $Z$ and $\alpha$ derived from the Euler-Lagrange equation reads
\begin{align}
\frac{m \ddot{Z}}{c^2} = \frac{2 M \dot{\alpha}}{\hbar} + m \nabla^2_{\Vec{r}} Z, \qquad
\frac{\ddot{\alpha}}{m c^2} = - \frac{2 M \dot{Z}}{\hbar} + \frac{\nabla^2_{\Vec{r}} \alpha}{m}. \label{eq-domain-dynamics}
\end{align}
In the massless limit $M \to 0$, the dynamics of $Z$ and $\alpha$ are independent of each other, giving 
linear dispersions:
\begin{align}
\omega = \pm c |\Vec{k}|, \label{eq-dispersion-rel}
\end{align}
with the frequencies $\omega$ both for $Z$ and $\alpha$, and the wave-number $\Vec{k} = (k_x, k_y)$.
Waves for $Z$ and $\alpha$ independently propagate as a ripplon and a localized magnon in the vicinity of the domain wall.

\begin{figure}[tbh]
\centering
\includegraphics[width=0.5\linewidth]{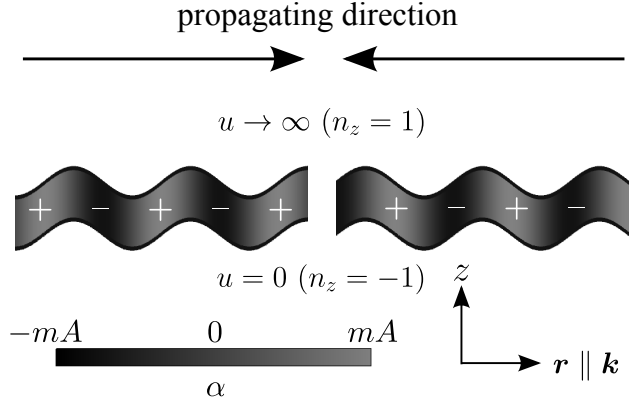}
\caption{\label{fig-ripplon} Schematic pictures of coupled ripplons and their propagating directions for solutions $(Z_1^+, \alpha_1^+)$ (left) and $(Z_1^-, \alpha_1^-)$ (right).
The middle shaded area show the region of the domain wall $|u| \approx 1$ ($n_z \approx 0$) and its tone shows the phase $\alpha$ of $u$ (direction of $(n_x,n_y)$).
The $+$ and $-$ signs show the areas for $\alpha > 0$ and $\alpha < 0$ respectively.
The vertical axis is the $z$--axis and the horizontal axis shows the direction of the wave vector $\Vec{k}$ in the $xy$--plane.
For left (right) figures, the coupled ripplon propagates in the right (left) direction.
See Ancillary files for animations of their dynamics.
}
\end{figure}

As long as $M \neq 0$, the dynamics of $Z$ and $\alpha$ couple to each other.
There are four typical solutions of Eq \eqref{eq-domain-dynamics}:
\begin{subequations}
\begin{align}
& Z_1^\pm = \frac{A_1^\pm}{m} \sin(\Vec{k} \cdot \Vec{r} \mp \omega_1 t + \delta_1^\pm), \quad
\alpha_1^\pm = \pm A_1^\pm \cos(\Vec{k} \cdot \Vec{r} \mp \omega_1 t + \delta_1^\pm), \label{eq-ripplon-NG} \\
& Z_2^\pm = \frac{A_2^\pm}{m} \sin(\Vec{k} \cdot \Vec{r} \pm \omega_2 t + \delta_2^\pm), \quad
\alpha_2^\pm = \pm A_2^\pm \cos(\Vec{k} \cdot \Vec{r} \pm \omega_2 t + \delta_2^\pm), \label{eq-ripplon-gap}
\end{align} \label{eq-ripplon-solution}
\end{subequations}
where $A_{1,2}^\pm \in \mathbb{R}$ and $\delta_{1,2}^\pm \in \mathbb{R}$ are arbitrary constants.
Waves of $Z$ and $\alpha$ couple to each other and propagate as a coupled ripplon with dispersions
\begin{align}
\begin{split}
& \omega_1 = \frac{\sqrt{M^2 c^4 + \hbar^2 c^2 \Vec{k}^2} - M c^2}{\hbar} = \frac{\hbar \Vec{k}^2}{2 M} + O(\Vec{k}^4), \\
& \omega_2 = \frac{\sqrt{M^2 c^4 + \hbar^2 c^2 \Vec{k}^2} + M c^2}{\hbar} = \frac{2 M c^2}{\hbar} + \frac{\hbar \Vec{k}^2}{2 M} + O(\Vec{k}^4). \label{eq-dispersion-general}
\end{split}
\end{align}
For the solutions of $(Z_1^\pm, \alpha_1^\pm)$, the coupled ripplons propagate in the direction parallel (for $+$ sign) and anti-parallel (for $-$ sign) to $\Vec{k}$ with a gapless quadratic dispersion $\omega_1$ showing type-II NG modes 
\cite{typeI-II-AB}. 
Figure \ref{fig-ripplon} shows the schematic pictures of coupled ripplons for $(Z_1^+, \alpha_1^+)$ (left) and $(Z_1^-, \alpha_1^-)$ (right).
In contrast to a quantized vortex in superfluids 
in which a Kelvin-wave is a combination of two translational modes in real space, 
the coupled ripplon is a combination of the translational mode in real space and the phase mode of the internal degree of freedom.
For the solutions of $(Z_2^\pm, \alpha_2^\pm)$, on the other hand, the coupled ripplons propagate in the opposite directions to $(Z_1^\pm, \alpha_1^\pm)$, respectively, with a gapped dispersion $\omega_2$, and do not behave as NG modes.
In the non-relativistic limit $c \to \infty$, the gap in $\omega_2$ diverges and the solutions $(Z_2^\pm, \alpha_2^\pm)$ disappear, and only $(Z_1^\pm, \alpha_1^\pm)$ for NG modes remain as solutions.
In the massless limit $M \to 0$, the gap in $\omega_2$ disappears and two dispersions $\omega_{1,2}$ become the same linear one: $\omega = c |k|$.

\section{Linear response theory}
Dynamics of a ripplon can also be analyzed by the linear response theory.
We consider the ansatz as the static domain-wall solution and its fluctuation: $u = u_0 + \delta u = u_0 + a_+ e^{i (\Vec{k} \cdot \Vec{r} - \omega t)} + a_-^\ast e^{- i (\Vec{k} \cdot \Vec{r} - \omega t)}$.
Inserting this ansatz into the dynamical equation \eqref{eq-dynamics}, we can obtain the Bogoliubov-de Gennes equation:
\begin{align}
\bigg(\frac{\omega^2}{c^2} \pm \frac{2 M \omega}{\hbar} \bigg) a_\pm = \bigg\{ (\Vec{k}^2 - \partial_z^2) + \frac{4 m e^{2 m z} \partial_z - m^2 ( 3 e^{2 m z} - 1 )}{1 + e^{2 m z}} \bigg\} a_\pm + O(a_\pm^2), \label{eq-Bogoliubov}
\end{align}
up to the linear order of $a_\pm$.
The normalizable solution is $a_\pm \propto e^{m z}$ and corresponding $\omega$ takes the value $\omega_{1,2}$ shown in Eq. \eqref{eq-dispersion-general}.
In the massless limit $M \to 0$, $\omega_1 = \omega_2 = c |\Vec{k}|$ gives the general solution $\delta u = e^{m z} \{ g_1^+ e^{i (\Vec{k} \cdot \Vec{r} - c |\Vec{k}| t)} + g_1^- e^{- i (\Vec{k} \cdot \Vec{r} - c |\Vec{k}| t)} + g_2^+ e^{i (\Vec{k} \cdot \Vec{r} + c |\Vec{k}| t)} + g_2^- e^{- i (\Vec{k} \cdot \Vec{r} + c |\Vec{k}| t)} \}$ with arbitrary constants $g_{1,2}^\pm \in \mathbb{C}$.
The localized magnon is obtained by taking $g_{1}^+ = - g_{1}^- = g_0 e^{i \delta}$ and $g_{2}^\pm = 0$ (parallel direction to $\Vec{k}$), or $g_{1}^\pm = 0$ and $g_{2}^+ = - g_{2}^- = g_0 e^{i \delta}$ (anti-parallel direction to $\Vec{k}$) with $g_0,\delta \in \mathbb{R}$.
The ripplon is obtained by taking $g_{1} = g_{2}^\ast = g_0 e^{i \delta}$ and $g_{3} = g_{4} = 0$ (parallel direction to $\Vec{k}$), or $g_{1} = g_{2} = 0$ and $g_{3} = g_{4}^\ast = g_0 e^{i \delta}$ (anti-parallel direction to $\Vec{k}$).
For $M \neq 0$ case, the solution is $\delta u_1 = e^{m z} \{ g_{1}^+ e^{i (\Vec{k} \cdot \Vec{r} - \omega_1 t)} + g_{1}^- e^{- i (\Vec{k} \cdot \Vec{r} + \omega_1 t)} \}$ and $\delta u_2 = e^{m z} \{ g_{2}^+ e^{i (\Vec{k} \cdot \Vec{r} + \omega_2 t)} + g_{2}^- e^{- i (\Vec{k} \cdot \Vec{r} - \omega_2 t)} \}$ with arbitrary constants $g_{1,2}^\pm \in \mathbb{C}$.
$g_{1,2}^+ = i A_{1,2}^+ e^{i \delta_{1,2}^+}$ and $g_{1,2}^- = 0$ ($g_{1,2}^+ = 0$ and $g_{1,2}^- = - i A_{1,2} e^{i \delta_{1,2}^-}$) correspond to the coupled ripplon solution $(Z_{1,2}^+, \alpha_{1,2}^+)$ ($(Z_{1,2}^-, \alpha_{1,2}^-)$) in Eq. \eqref{eq-ripplon-solution}.

\section{Commutation relation}

We obtain gapless and localized magnon and ripplon with linear dispersions, Eq.~\eqref{eq-dispersion-rel}, only in the massless limit $M \to 0$ and the coupled ripplon with quadratic dispersion, Eq.~\eqref{eq-ripplon-NG}, with $M \neq 0$ in the vicinity of the domain wall.
These modes are type-I (for $M \to 0$) and type-II (for $M > 0$) NG modes as a consequence of the spontaneous breaking of the $U(1)$ symmetry for the phase rotation and the translational symmetry: $H_1 \simeq U(1) \times \mathbb{R}^3 \to H_2 \simeq \mathbb{R}^2$ under the appearance of the domain wall.
In both cases, our dispersions saturate the equality of the Nielsen-Chadha inequality \cite{Nielsen:1975hm}: $N\sub{I} + 2 N\sub{II} \geq N\sub{BG}$, where $N\sub{I}$, $N\sub{II}$, and $N\sub{BG}$ are the total numbers of the type-I NG modes, the type-II NG modes, and broken generators (BG) which correspond to spontaneously broken symmetries.
Recently, it has been shown 
in Refs.~\cite{Watanabe:2012hr,Hidaka:2012ym} 
that for internal symmetry
the equality of the Nielsen-Chadha inequality is saturated as the Watanabe-Brauner's relation \cite{Watanabe:2011ec}:
\begin{align}
N\sub{BG} - N\sub{NG} = \frac{1}{2} \mathrm{rank} \rho, \quad
\rho_{i,j} = \lim_{V \to \infty} \frac{1}{V} \int d^3\Vec{x}\: (- i [\Omega_i, \Omega_j]) 
 \Big|_{u = u_0}, \label{eq-Watanabe-Brauner}
\end{align}
where $N\sub{NG} = N\sub{I} + N\sub{II}$ is the number of NG modes, $V$ is the volume of the system, 
$\Omega_i$ is the Noether's 
charge or a generator of broken symmetries, 
and $[\cdot,\cdot]$ is 
a commutator or the Poisson bracket in classical level.
According to this relation, $N\sub{BG} \neq N\sub{NG}$ takes place when 
commutators of  
broken generators are non-vanishing.
This relation has been proven for internal symmetries 
such as the Heisenberg ferromagnet 
and has been 
confirmed for space-time symmetries 
such as a quantized vortex in superfluid 
and a two-dimensional skyrmion in ferromagnets 
\cite{Watanabe:2014pea}. 
In our case, on the other hand, broken generators consist of  one internal symmetry and one spatial symmetry which intuitively commute because underlying symmetries are 
the direct product and are independent of each other, {\it i.e.}, $H_1 / H_2 \simeq U(1) \times \mathbb{R}$.
To check whether the relation, 
Eq.~\eqref{eq-Watanabe-Brauner}, also holds in our case or not, we directly calculate the commutation relation between symmetry generators of the internal $U(1)$ phase rotation and the translation.
Defining the  momenta $v$ conjugate to $u$ as
\begin{align}
v = \frac{\partial \mathcal{L}\sub{G}}{\partial \dot{u}} = \frac{\dot{u}^\ast}{c^2 (1 + |u|^2)^2} + \frac{i M u^\ast}{\hbar (1 + |u|^2)}, \label{eq-momentum}
\end{align}
the Noether's charges for the phase rotation and the translation along $z$-axis are obtained as 
\begin{align}
& \Theta =  \int dz\: J^0_\alpha , \quad J^0_\alpha = i u v, \label{eq-phase-operator} \\ 
& P = \int dz\: J^0_Z , \quad J^0_Z = (\partial_z u) v \label{eq-translation-operator},
\end{align}
respectively. 
The commutator between $P$ and $\Theta$
can be calculated from 
$[u(z_1),v(z_2)] = i \delta(z_1-z_2)$, to yield 
\begin{align}
\begin{split}
 [P,\Theta] &= \int dz_1\: \int dz_2\: [J^0_Z(z_1), J^0_\alpha(z_2)] \\
&= i \int dz_1\: \int dz_2\: [(\partial_{z_1} u(z_1)) v(z_1), u(z_2) v(z_2)] \\
&= i \int dz_1\: \int dz_2\: \{ (\partial_{z_1} u(z_1)) [v(z_1), u(z_2)] v(z_2) + u(z_2) \partial_{z_1} [u(z_1), v(z_2)] v(z_1) \} \\
&= \int dz_1\: \int dz_2\: \{ (\partial_{z_1} u(z_1)) v(z_2) + u(z_2) (\partial_{z_1} v(z_1)) \} \delta(z_1 - z_2) \\
&= \int dz\: \partial_z ( u(z) v(z) ). 
\end{split}
\end{align}

For the static domain-wall solution, the first term in Eq. \eqref{eq-momentum} does not contribute to the commutator because of $\dot{u} = 0$.
Consequently, the commutator becomes
\begin{align}
\begin{split}
- i [P, \Theta] &= 
\frac{M}{\hbar} \int dz\: \partial_z \bigg( \frac{|u|^2}{1 + |u|^2} \bigg) 
= \frac{M}{\hbar} \left[ \frac{|u|^2}{1 + |u|^2}\right]^{z= +\infty}_{z=-\infty}
= \frac{2 M}{\hbar} \bigg( \frac{1}{2} \left[ 1- n_z\right]^{z= +\infty}_{z=-\infty} \bigg) \\
&\equiv \frac{2 M W}{\hbar}.
\label{eq:W}
\end{split}
\end{align}
$W$ is precisely the topological charge of the domain wall 
and is proportional to the tension of the domain wall 
\cite{Abraham:1992vb,Arai:2002xa} 
(see Appendix \ref{sec:topological}).
Evaluating this in the domain wall background 
$u=u_0$, we find $W = 1$.
As a result, two generators $P$ and $\Theta$ do not commute as long as $M \neq 0$, giving $N\sub{BG} - N\sub{NG} = 1$ 
and one type II NG mode, or commute in the massless limit $M \to 0$ giving two type I NG modes, which is consistent with our result.

\section{Conclusion}
In conclusion, we have considered NG modes excited on one flat domain wall in the $\mathbb{C}P^1$ models with the Ising potential.
NG modes in the relativistic model are the localized magnon for the $U(1)$ phase rotation and the translational ripplon which are independent of each other and have linear dispersions.
In the non-relativistic limit, on the other hand, there is one coupled ripplon with a quadratic dispersion as the combination of the localized magnon and the ripplon.
We also find the coupled localized magnon and the ripplon in the interpolating model connecting 
the relativistic and non-relativistic theories 
even though it has the Lorentz invariance.
The numbers of NG modes saturate the equality of the Nielsen-Chadha inequality, 
and also satisfy the Watanabe-Brauner's relation in which the commutator between two generators of the internal phase mode and spatial translational mode gives 
the topological domain wall charge.

Quantum effects on localized type-II NG modes 
remain as an important problem, which was studied for 
a vortex with non-Abelian localized modes \cite{Nitta:2013wca}.

The term $|u|^2 / (1 + |u|^2) = (1/2) (1- n_z)$ 
in Eq.(\ref{eq:W}) is
known as the momentum map in symplectic geometry 
and the D-term in supersymmetric gauge theory. 
Therefore, 
our model can be extended to the ${\mathbb C}P^n$ model, Grassmann sigma model \cite{Isozumi:2004jc,Eto:2006pg},
sigma models on more general K\"ahler target manifolds, 
and non-Abelian gauge theories. 
A domain wall in two-component Bose-Einstein condensates 
has a different structure of NG modes  
\cite{Takeuchi:2013mwa}, 
although there are also translational and internal $U(1)$ zero modes \cite{footnote3}. 
This may be because the $U(1)$ zero mode 
is non-normalizable in their case. 
If one couples a gauge field, their model reduces to 
ours in strong gauge coupling limit (see Appendix~\ref{sec:rel-nrel}), with the internal $U(1)$ mode
becoming normalizable.

\section*{Acknowledgment}
We thank H. Takeuchi for useful discussions 
and H. Watanabe for explaining their results.
We also thank the anonymous referees for the helpful suggestions and comments.
This work is supported in part by Grant-in-Aid for Scientific Research (Grants No. 26870295 (M.K.) and No. 25400268 (M.N.)) and the work of M. N. is also supported in part by the ``Topological Quantum Phenomena" Grant-in-Aid for Scientific Research on Innovative Areas (No. 25103720) from the Ministry of Education, Culture, Sports, Science and Technology (MEXT) of Japan.

\appendix
%%%%%%%%%%%%%%%%%%%%%%%%%
\section{Derivation of Lagrangian density interpolating between relativistic and non-relativistic models}\label{sec:rel-nrel}

We consider a $U(1)$ gauge theory coupled with two charged complex scalar fields $\phi = (\phi_{1}, \phi_{2})^T$: 
\begin{align}
\mathcal{L}\sub{g} = - \frac{1}{4 e^2} F_{\mu\nu} F^{\mu\nu} + |D_\mu \phi|^2 - \frac{e^2}{2} (\phi^\dagger \phi - \phi_0^2)^2 - \frac{m^2}{4 \phi_0^2} \{ \phi_0^4 - (\phi^\dagger \sigma_z \phi)^2 \},
\end{align}
with the gauge coupling $e$, 
the vacuum expectation value $\phi_0$  
and $D_\mu = \partial_{\mu} - i A_{\mu}$.
The last term stabilizes the $\phi^\dagger \sigma_z \phi = \pm \phi_0^2$ as vacua, where $\sigma_z = \mathrm{diag}(1,-1)$ is the third Pauli matrix.
With separating the total energy into the invariant mass and the others: $\phi \to e^{- i M c^2 t / \hbar} \phi$ with the particle mass $M$, $\mathcal{L}\sub{g}$ becomes
\begin{align}
\begin{split}
\mathcal{L}\sub{g} - \frac{M^2 c^2 \phi^\dagger \phi}{\hbar^2} &= - \frac{1}{4 e^2} F_{\mu\nu} F^{\mu\nu} + |D_\mu \phi|^2 + \frac{i M \{ \phi^\dagger (D_t \phi) - (D_t \phi)^\dagger \phi \}}{\hbar} - \frac{e^2}{2} (\phi^\dagger \phi - \phi_0^2)^2 \\
&\quad\:  - \frac{m^2}{4 \phi_0^2} \{ \phi_0^4 - (\phi^\dagger \sigma_z \phi)^2 \}.
\end{split}
\end{align}
$\mathcal{L}\sub{g}$ is invariant under the following Lorentz transformation
\begin{align}
\begin{split}
& t^\prime = \gamma \bigg( t - \frac{\Vec{v} \cdot \Vec{x}}{c^2} \bigg), \quad
\Vec{x}^\prime = \gamma (\Vec{x} - \Vec{v} t), \quad
A_0^\prime = \gamma(A_0 + \Vec{v} \cdot \Vec{A}), \quad
\Vec{A}^\prime = \gamma \bigg( \frac{\Vec{v}}{c^2} A_0 + \Vec{A} \bigg) \\
& \phi^\prime = e^{i S} \phi, \quad
S = - \frac{(1 - \gamma) M c^2 t}{\hbar} - \frac{\gamma M \Vec{v} \cdot \Vec{x}}{\hbar}, \quad
\gamma = \frac{1}{\sqrt{1 - v^2 / c}},
\end{split}
\end{align}
showing the Lorentz symmetry of $\mathcal{L}\sub{g}$.
The Lorentz symmetry reduces to Galilean or Schr\"odinger symmetry in the nonrelativistic limit $c \to \infty$.
By taking the strong coupling limit $e \to \infty$, 
the model reduces to the $\mathbb{C}P^1$ model  
[see 
A. D'Adda, M. Luscher, and P. Di Vecchia, Nucl.\ Phys.\ B {\bf 146}, 63 (1978)].
By rewriting $\phi = \phi_0 (1, u) / \sqrt{1 + |u|^2}$ with complex projective coordinate $u$, we obtain
\begin{align}
\frac{\mathcal{L}\sub{g}}{\phi_0^2} - \frac{M^2 c^2}{\hbar^2} &= \frac{|\dot{u}|^2}{c^2 (1 + |u|^2)^2} - \frac{|\nabla u|^2}{(1 + |u|^2)^2} + \frac{i M ( u^\ast \dot u - \dot{u}^\ast u )}{\hbar (1 + |u|^2)} - \frac{m^2 |u|^2}{(1 + |u|^2)^2}.
\end{align}
We obtain $\mathcal{L}\sub{G}$ in Eq.~(4) in the main text as $\mathcal{L}\sub{G} = \mathcal{L}\sub{g} - M^2 c^2 \phi_0^2 / \hbar^2$.
$\mathcal{L}\sub{G}$ is invariant under the following Lorentz transformation
\begin{align}
\begin{split}
& t^\prime = \gamma \bigg( t - \frac{\Vec{v} \cdot \Vec{x}}{c^2} \bigg), \quad
\Vec{x}^\prime = \gamma (\Vec{x} - \Vec{v} t), \quad
u^\prime = e^{i \mathcal{S}} u, \\
& \partial_t \mathcal{S} = - \frac{(1 - \gamma) M c^2 (1 + |u|^2)}{\hbar}, \quad
\nabla \mathcal{S} = - \frac{\gamma M \Vec{v} (1 + |u|^2)}{\hbar},
\end{split}
\end{align}
showing the Lorentz symmetry of $\mathcal{L}\sub{G}$.
The Lorentz symmetry also reduces to 
the Galilean or Schr\"odinger symmetry in the nonrelativistic limit $c \to \infty$.

%%%%%%%%%%%%%%%%%%%%%
\section{Topological charge of a domain wall}\label{sec:topological}
Here, we describe 
the topological charge of a static and flat domain wall 
interpolating between two discrete vacua $u=0$ and $u=\infty$.
The Bogomol'nyi completion for the tension (energy 
per unit length) of a static domain wall can be obtained as
\beq
E = \int d z 
\frac{|\partial_{z} u| +m^2|u|^2 }
{(1+|u|^{2})^{2}}
= \int d z 
\frac{
|\partial_{z} u \mp mu|^2 
\pm m (u^{\ast} \partial_z u + u \partial_z u^{\ast})}
{(1+|u|^{2})^{2}}
\geq
|T| , \label{eq:BPS-bound-wall}
\eeq
where $\partial_z$ denotes 
the differentiation with respect to $z$,  
and $T$ is 
the topological charge of the domain wall: 
\begin{eqnarray}
T &\equiv& m \int  d z \frac{ u^{\ast} \partial_z u 
+ u \partial_z u^{\ast}}{(1+|u|^{2})^{2}}   
= {m\over 2} \int  d z \partial_z \left( {1-|u|^2 \over 1+|u|^2}\right)
= {m\over 2} \left[ {1-|u|^2 \over 1+|u|^2} \right]^{z=+\infty}_{z=-\infty} \nonumber\\
&=& {m\over 2} \left[ n_z \right]^{z=+\infty}_{z=-\infty}
= m \left[(1-W)\right]^{z=+\infty}_{z=-\infty}. 
\end{eqnarray}
With a fixed boundary condition 
and the topological charge $T$, 
the most stable configurations 
saturate the inequality in Eq.~(\ref{eq:BPS-bound-wall}) 
and satisfy the BPS equation  
\begin{equation}
\partial_z u \mp m u =0,  
\end{equation}
obtained by $|...|^2=0$ in  Eq.~(\ref{eq:BPS-bound-wall}). 
This BPS equation can be immediately solved as 
%\cite{Abraham:1992vb, Arai:2002xa}
(see Refs.~[12,14] in the main text)
\beq
 u_0 = e^{\pm m (z - Z) + i \alpha} , \label{eq:wall}
\eeq
where $\pm$ denotes a domain wall and an anti-domain wall,
with the tension 
\beq 
 |T| = m,
\eeq


%%%%%%%%%%%%%%%%%%%%%%%%%%

\end{document}